\documentclass[%
 reprint,
 amsmath,amssymb,
 aps,
]{revtex4-2}
\usepackage[english]{babel}
\usepackage{graphicx}
\usepackage{dcolumn}
\usepackage{bm}
\usepackage{amssymb}
\usepackage{amsfonts}
\usepackage[sort&compress]{natbib}
\usepackage{color}

\usepackage{xcolor}

\begin{document}
\preprint{APS/123-QED}
\title{On kinetic energy localization in fluid flow}
\author{Damian Śnieżek}
\email{damian.sniezek@uwr.edu.pl}

\affiliation{%
 Institute of Theoretical Physics, University of Wroc{\l}aw, pl. M. Borna 9, 50-204 Wroc{\l}aw, Poland
}%

\date{\today}

\begin{abstract}
This works focuses on participation number -- a parameter that allows to quantitatively asses the level of kinetic energy localization. The author presents a clear way of deriving participation number in a continuous case without making any assumptions about the system, fluid or flow regime. 
Moreover, a method of computing participation number in discretized cases is discussed and verified against well known analytical solutions using three methods, in which one was used previously in research on fluid flow through porous media. A robust formula, that works for both uniform and nonuniform discretization grids is presented.
\end{abstract}
\maketitle

There are many fields that can benefit from kinetic energy analysis. It was used to study and optimize the nutrient transport in hydroponic system \cite{guzman2019turbulent}. It provides more fundamental understanding of blood flow in human aorta \cite{buchner2024analysis}. It can be used to optimize placement of hydrokinetic devices along rivers \cite{walsh2012importance}.
In analysis of fluid flow through system of any kind, it is important to use the proper tools and methods dedicated to the kind of the flow one works with.
There are few important, general aspects that allow to describe the system, such as geometry, flow regime or the type of fluid. Sometimes, in specific cases, it can be hard to decide what tools should be used. A more in-depth analysis requires tools that give unambiguous information about the flow. That is why having robust tools, that work as wide range of cases as possible is important.

One of the crucial information, that brings more understanding of complex flow dynamics in various fields, is the spatial distribution of kinetic energy, or in another words, level of its localization. It tells whether the whole system participates equally in the transport or if only some parts of the system are important.
In recent years, analysis of kinetic energy distribution plays more and more important role. In \cite{andrade1999inertial}, in order to analyze  the weakening of channeling effect in inertial flow regime in fluid flow through disordered porous media, the authors introduced participation number, a parameter that allows to quantify kinetic energy localization.
It was modified by Hyman to describe channeling in fracture networks \cite{hyman2020flow}.
It was also used in \cite{seybold2021localization}, where the authors studied how to control certain aspects of flow structure in fluid flow through porous medium.
The concept of participation number has proved itself useful in describing transitions between flow regimes \cite{andrade1999inertial, sniezek2024inertia}.
Up to this point, little attention was paid to the derivation of participation number itself. The aim of this work is to clearly present the idea of participation number as a parameter that quantifies localization and to verify its discretized form based on well known analytical solutions.

The concept of participation number $\pi$, as a parameter that quantitatively describes the level of kinetic energy localization is based on \cite{bonifacio1994spectrum}. It was introduced to the field of fluid flow through porous media in \cite{andrade1999inertial}, where it was defined as
\begin{equation}
    \label{discrete_pi_def}
    \pi \equiv \left( n \sum_{i=1}^n q_i^2 \right)^{-1}, ~~~~\pi \in \left[\frac{1}{n}, 1 \right],
\end{equation}
where $n$ is the number of discretization grid cells. Parameter $q_i \equiv e_i / \sum_{i=j}^n e_j$ has the meaning of fraction of total kinetic energy contained in $i$-th discretization cell \cite{andrade1999inertial}. The limits for possible values of $\pi$ come from two opposite cases. $\pi = 1/n$ when the whole kinetic energy is contained in a single cell. On the other hand, $\pi = 1$ when kinetic energy in uniformly distributed in the system, i.e. $q_i = 1 / n$ for $i = 1, 2, \dots, n$. Moreover, it can be noted that the factor $n^{-1}$ on the right hand side of equation \eqref{discrete_pi_def} normalizes the set of possible values, so that in a limit where $n \rightarrow \infty$, $\pi \in [0,1]$. 

The above approach assumes that the velocity field is discretized into a uniform grid with each cell having the same shape and volume. It can be viewed as a special case of a more general formula which definition is hereby presented for a continuous case without making any assumptions about type of fluid, geometry of the system of flow regime. 

We define the fractional concentration of total kinetic energy as a function of position in space, $q = q(x, y, z)$. It can be expressed as
\begin{equation}
    \label{continous_q_def}
    q(x, y, z) = \frac{\rho(x, y, z) |u|^2 (x, y, z)}{2 E_\mathrm{tot}}
\end{equation}
where $E_\mathrm{tot} = \int_V\frac{\rho(x, y, z)u^2(x ,y, z)}{2}dV$ is the total kinetic energy in the whole system, $\rho = \rho(x,y,z)$ is fluid's mass density and $u = u(x,y,z)$ is the velocity field. As can be noted, $q$ has dimensions of $\mathrm{volume}^{-1}$.

In order to find limits and normalization factor for participation number in a general case one has to integrate $q^2(x,y,z)$ over the volume of considered system and consider two opposite cases of kinetic energy distribution. In the case of maximum localization, the whole kinetic energy is contained in a single point $(x_0, y_0, z_0)$. Kinetic energy can be defined using Dirac delta as $E_k(x,y,z) = E_\mathrm{tot}\delta(x-x_0, y-y_0, z-z_0)$. According to this definition of kinetic energy, parameter $q$ is defined as $q(x,y,z) = \delta(x-x_0, y-y_0, z-z_0)$. Therefore
\begin{multline}
    \label{q^2_max_localized}
    \int_V q^2(x,y,z)~dxdydz \\
    = \int_V \delta^2(x-x_0, y-y_0, z-z_0)~dxdydz
    \propto\frac{1}{\epsilon},
\end{multline}
where $\epsilon$ is an arbitrary small number that emerges from regularization of Dirac delta with a narrow Gaussian. In the case of uniform energy distribution, the kinetic energy can be defined as $E_k(x,y,z) = E_\mathrm{tot} / V$, thus $q(x,y,z) = 1 / V$. In this case
\begin{multline}
    \label{q^2_uniform}
    \int_V q^2(x,y,z)~dxdydz
    = \int_V V^{-2}~dV' \\
    = V^{-2} \int_V~dV'
    = \frac{1}{V},
\end{multline}
where system volume $V$ is treated as a constant. From the fact that
\begin{equation}
    \label{int_q^2_inverse_bounds}
    \left(\int_V q^2(x,y,z)~dxdydz\right)^{-1} \in \left[\epsilon, V\right],
\end{equation}
it is clear that a correct normalization factor in this case is inverse of the volume of considered system $V^{-1}$. On the other hand volume integral over $q^2$ has dimensions of $\mathrm{Volume}^{-1}$, which if multiplied by system's volume, becomes a dimensionless quantity.
This shows, that the participation number in a general, continuous case is defined as
\begin{equation}
    \label{pi_continuous_def}
    \pi \equiv \left(V \int_V q^2(x,y,z)~dxdydz\right)^{-1}, ~~~~\pi \in \left[\frac{\epsilon}{V}, 1\right],
\end{equation}
where $\epsilon$ is an arbitrary small number. In a limit where $\epsilon \rightarrow 0$, $\pi \in [0, 1]$. The same definition of participation number was given in \cite{seybold2021localization} without derivation.

It is also possible to obtain a general definition of participation number in nonuniformly discretized system. The discretized form of the definition \eqref{continous_q_def} is
\begin{equation}
    \label{discrete_general_q_def}
    q_i = \frac{\rho_i u_i^2}{2 E_\mathrm{tot}},
\end{equation}
where $i$ is the index denominating a cell of a nonuniform grid and $E_\mathrm{tot} = \sum_{i=1}^n \frac{1}{2} \rho_i u_i^2 V_i$.
The next step is to consider the same opposite kinetic energy distributions as in continuous case.
For maximum localization the whole kinetic energy is contained inside one grid cell, thus the kinetic energy takes the form
\begin{equation}
    \label{E_non_uniform_max_localized}
    E_i = \begin{cases}
        \frac{1}{2} \rho_i u_i^2 V_i & \text{if } i = k,\\
        0 & \text{if } i \neq k
    \end{cases},
\end{equation}
where $k$ is an arbitrary grid cell that contains the whole kinetic energy. It can be noted that $E_\mathrm{tot} = \sum_{i=1}^n \frac{1}{2} \rho_i u_i^2 V_i = \frac{1}{2} \rho_k u_k^2 V_k$. In this case
\begin{equation}
    \label{q_non_uniform_max_localized}
    q_i = \begin{cases}
        \frac{\rho_i u_i^2}{2 E_\mathrm{tot}} = \frac{1}{V_i} & \text{if } i = k,\\
        0 & \text{if } i \neq k
    \end{cases}
\end{equation}
leads to the following conclusion
\begin{equation}
    \label{sum_q_non_uniform_max_localized}
    \sum_{i=1}^n q^2_i V_i 
    = \sum_{i=1}^n \frac{(\rho_i u_i)^2}{4 E_\mathrm{tot}^2} V_i 
    = \frac{(\rho_k u_k)^2}{(\rho_k u_k V_k)^2} V_k
    = \frac{1}{V_k}.  
\end{equation}

On the other hand, when kinetic energy is uniformly distributed throughout the system, i.e.
\begin{equation}
    \label{E_non_uniform_uniform_distribution}
    E_i = 
    \frac{V_i}{V} \sum_{j=1}^n \frac{1}{2} \rho_j u^2_j V_j =
    V_i\frac{E_\mathrm{tot}}{V},
\end{equation}

\begin{equation}
    \label{q_non_uniform_uniform_distribution}
    q_i = 
    \frac{\rho_i u_i^2}{2 E_\mathrm{tot}} = \frac{\rho u^2}{2 E_\mathrm{tot}},
\end{equation}
where $V$ is the volume of considered system, $\rho_i = \rho$ and $u_i = u$ for all $i = 1,2, \dots, n$ . In this case
\begin{equation}
    \label{sum_q_non_uniform_uniform_distribution}
    \sum_{i=1}^n q^2_i V_i
    = \sum_{i=1}^n \left(\frac{\rho u^2}{2 E_\mathrm{tot} }\right)^2  V_i 
    = \left(\frac{\rho u^2}{2 E_\mathrm{tot} }\right)^2  V
\end{equation}
has dimensions of $\mathrm{Volume}^{-1}$. Similarly as in \eqref{int_q^2_inverse_bounds}, those two cases define the boundaries of possible values
\begin{equation}
    \left(\sum_{i=1}^n q^2_i V_i\right)^{-1} \in \left[V_k, \frac{4 E_\mathrm{tot}^2}{\rho^2 u^4 V}\right]
\end{equation}
In order to obtain a dimensionless quantity from \eqref{sum_q_non_uniform_max_localized} and \eqref{sum_q_non_uniform_uniform_distribution} we multiply both equations by volume of considered system $V$. It also serves as the normalization constant for nonuniform discretization grid. Thus the definition of participation number takes the form
\begin{equation}
    \label{pi_discretized_non_uniform_def}
    \pi \equiv \left(V \sum_{i=1}^n q_i^2V_i \right)^{-1} \in \left[\frac{V_k}{V}, \frac{4 E_\mathrm{tot}^2}{\rho^2 u^4 V^2}\right].
\end{equation}
Remembering that the upper limit comes from uniform energy distribution it is clear that $\frac{4 E_\mathrm{tot}^2}{\rho^2 u^4 V^2} = 1$. Since $V_k$ in the lower limit is an arbitrary grid cell the ratio $\frac{V_k}{V} \rightarrow 0$ as $n \rightarrow \infty$. Thus, is such a limiting case
\begin{equation}
    \label{pi_nonuniform_grid_limits}
    \pi \in [0, 1].
\end{equation}

Two systems with known analytical solutions for velocity field were considered; Couette flow and fluid flow through cylindrical pipe \cite{chorin1990mathematical}. In each case participation numbers were computed analytically. Then the results were used as a reference to verify whether formulae \eqref{discrete_pi_def} and \eqref{pi_discretized_non_uniform_def} give correct numerical results in case of uniform and nonuniform mesh grid. Numerical simulations were performed with OpenFOAM 12 \cite{jasak2007openfoam} using the finite volume method. The grids were made using standard meshing tools such as blockMesh and snappyHexMesh. Steady state simulations were made with incompressible solver. Fully converged solutions were obtained for all simulations.

In the case of two--dimensional Couette flow there are two parallel boundaries distant from each other by $h$. The upper boundary moves with velocity $U$ with respect to bottom plane in positive $x$ direction. The fluid is accelerated due to viscous forces acting between its layers. Velocity in Cartesian coordinates can be expressed as $\vec{u} = \left(U\frac{y}{h}, 0\right)$. To compute participation number one can investigate part of the system with arbitrary length $L$, so that $y \in [0,h]$ and $x \in [0,L]$. Surface area of such a system is $A = Lh$. Total kinetic energy can be found by integrating
\begin{equation}
    \label{couette_tot_kin_energy}
    E_\mathrm{tot} = \int_0^L\int_0^h \frac{\rho u^2}{2} ~dydx = \frac{\rho L U^2 h}{6}.
\end{equation}
In this system 
\begin{equation}
    \label{q_couette}
    q \equiv \frac{\rho u^2}{2E_\mathrm{tot}} = \frac{3y^2}{Lh^3},
\end{equation}
thus
\begin{equation}
    \label{pi_couette}
    \pi = \frac{5}{9}
\end{equation}

    Participation number was computed according to equations \eqref{discrete_pi_def} and \eqref{pi_discretized_non_uniform_def} in cases of uniform and nonuniform meshes. The results are presented in Table \ref{tab:couette_uniform} and Table \ref{tab:couette_non_uniform} respectively. As can be seen, both methods produce results very close to the analytical result in case of uniform discretization. Maximum absolute difference is of the order $10^{-7}$.
    On the other hand, equation \eqref{discrete_pi_def} gives incorrect results in case of non uniform meshes, whereas equation \eqref{pi_discretized_non_uniform_def} gives the correct values.
\begin{table}[]
    \centering
    \begin{tabular}{|c|c|c|}
        \hline
         $L_c$ & $\pi$ & $\pi^*$\\
        \hline
         1/80  & 0.55520272 & 0.55520271 \\ 
         1/100 & 0.55484893 & 0.55484891 \\ 
         1/120 & 0.55443804 & 0.55443803 \\ 
         1/140 & 0.55400012 & 0.55400017 \\ 
         1/180 & 0.55216903 & 0.55216901 \\ 
         1/200 & 0.55060371 & 0.55060381 \\ 
        \hline
    \end{tabular}
    \caption{Values of participation number computed according to \eqref{discrete_pi_def} ($\pi$) and  \eqref{pi_discretized_non_uniform_def} ($\pi^*$) in case of 2D Couette flow calculated on different levels of uniform mesh refinement. $L_c$ means characteristic length, i.e. $L_c^i = V_i^{\frac{1}{3}}$}
    \label{tab:couette_uniform}
\end{table}

\begin{table}[]
    \centering
    \begin{tabular}{|c|c|c|c|}
        \hline
        $n$ & $\langle L_c \rangle$ & $\pi$ & $\pi^*$\\
        \hline
         1408 & 4.061721e-03 & 0.51434 & 0.55888 \\ 
         26128 & 1.496354e-03 & 0.45295 & 0.55531 \\ 
         33088 & 1.411611e-03 & 0.50503 & 0.55488 \\ 
         131188 & 8.913942e-04 & 0.50360 & 0.55098 \\ 
         124276 & 8.799376e-04 & 0.27031 & 0.55186 \\ 
        \hline
    \end{tabular}
    \caption{Values of participation number computed according to \eqref{discrete_pi_def} ($\pi$) and  \eqref{pi_discretized_non_uniform_def} ($\pi^*$) in case of 2D Couette flow calculated on different levels of nonuniform mesh refinement. The meshes differ by total number of cells, their size and volume of refinement regions. $\langle L_c\rangle$ -- average characteristic length, $n$ -- number of mesh cells.}
    \label{tab:couette_non_uniform}
\end{table}

Here, we assume fluid flow through cylindrical pipe with radius $a$. No slip conditions  were imposed on the wall. Fluid flow is driven by known pressure difference $\Delta p$. We investigate kinetic energy localization in a part of the system with length $L$ and volume $V=\pi a^2 L$. Velocity can be described in cylindrical coordinates as $u(r,\theta,z) = \left(0,0,\frac{\Delta p}{4\mu}(a^2-r^2)\right)$. In this case we have the total kinetic energy
\begin{equation}
    \label{pipe_tot_kin_energy}
    E_\mathrm{tot} = \int_0^L\int_0^{2\pi}\int_0^a \frac{\rho u^2}{2}~dz d\theta dr = \frac{1}{6}\pi L \rho \left(\frac{\Delta p}{4\mu}\right)^2 a^6.
\end{equation}
In this case
\begin{equation}
    \label{q_pipe}
    q =\frac{3(a^2 - r^2)^2}{\pi L a^6},
\end{equation}
therefore
\begin{equation}
    \label{pi_pipe}
    \pi = \frac{5}{9}.
\end{equation}

Participation number was computed according to equations \eqref{discrete_pi_def} and \eqref{pi_discretized_non_uniform_def} for uniform and non uniform discretization grids, similarly as for Couette flow. The results are presented in Table \ref{tab:hagen-poiseuille_uniform} and \ref{tab:hagen-poiseuille_non_uniform} respectively. It can be seen, that both formulas produce correct results for uniform meshes. On the other hand, only equation \ref{pi_discretized_non_uniform_def} gives correct results for nonuniform grids. 

\begin{table}[]
    \centering
    \begin{tabular}{|c|c|c|}
        \hline
        $n$ & $\pi$ & $\pi^*$\\
        \hline
            11200 & 0.547971175 & 0.54797118 \\
            88480 & 0.550236528 & 0.55023657 \\ 
            300720 & 0.55184979 & 0.55184987 \\ 
            707840 & 0.55275169 & 0.55275173 \\ 
            1377600 & 0.55725573 & 0.55725572 \\ 
        \hline
    \end{tabular}
    \caption{Values of participation number computed according to \eqref{discrete_pi_def} ($\pi$) and  \eqref{pi_discretized_non_uniform_def} ($\pi^*$) in case of 3D flow in cylindrical pipe calculated on different levels of uniform mesh refinement. The meshes differ by total number of cells $n$.}
    \label{tab:hagen-poiseuille_uniform}
\end{table}

\begin{table}[]
    \centering
    \begin{tabular}{|c|c|c|c|}
        \hline
        $n$ & $\langle L_c \rangle$ & $\pi$ & $\pi^*$\\
        \hline
         543600 & 0.047018 & 0.09605 & 0.55517 \\ 
         1321755 & 0.032092 & 0.01770 & 0.55492 \\ 
         183375 & 0.079808 & 0.53230 & 0.55776 \\ 
         10212792 & 0.012951 & 0.00525 & 0.55375 \\ 
        \hline
    \end{tabular}
    \caption{Values of participation number computed according to \eqref{discrete_pi_def} ($\pi$) and  \eqref{pi_discretized_non_uniform_def} ($\pi^*$) in case of 3D flow in cylindrical pipe calculated on different levels of nonuniform mesh refinement. The meshes differ by total number of cells, their size and volume of refinement regions. $\langle L_c\rangle$ -- average characteristic length, $n$ -- number of mesh cells.}
    \label{tab:hagen-poiseuille_non_uniform}
\end{table}

This work presents derivation of a general formula for computing participation number, a parameter that quantifies kinetic energy localization. There were no assumptions made about the type of fluid, or the flow regime, therefore presented approach works in all cases, including multiphase flow. Moreover, a robust formula, that works for all types of discretization grids, is derived. The method was verified and proved to work based on two well known analytical solutions of Couette flow and Hagen--Poiseuille flow in a tube with circular cross--section. Presented method, with little modification, could also be used do compute the normalized localization of any scalar field, not only kinetic energy.

\end{document}